\documentclass[prb,superscriptaddress,twocolumn,amsmath,amssymb,showpacs,floatfix]{revtex4-2}
\usepackage{multirow}
\usepackage{graphicx}
\usepackage{amsmath}
\usepackage{dcolumn}
\usepackage{bm}
\usepackage{color}
\usepackage{adjustbox}
\usepackage{siunitx}
\hfuzz=\maxdimen
\begin{document}
	
\title{Physical properties and first-principles calculations of an altermagnet candidate Cs$_{1-\delta}$V$_2$Te$_2$O}
	
\author{Chang-Chao Liu}
\thanks{These authors contributed equally.}
\affiliation{School of Physics, Zhejiang University, Hangzhou 310058, China}

\author{Jing Li}
\thanks{These authors contributed equally.}
\affiliation{School of Physics, Zhejiang University, Hangzhou 310058, China}

\author{Ji-Yong Liu}
\affiliation{Department of Chemistry, Zhejiang University, Hangzhou 310058, China}

\author{Jia-Yi Lu}
\affiliation{School of Physics, Zhejiang University, Hangzhou 310058, China}

\author{Hua-Xun Li}
\affiliation{School of Physics, Zhejiang University, Hangzhou 310058, China}

\author{Yi Liu}
\affiliation{Department of Applied Physics, Zhejiang University of Technology, Hangzhou 310023, China}

	
\author{Guang-Han Cao}
\email[corresponding author: ]{ghcao@zju.edu.cn}
\affiliation{School of Physics, Zhejiang University, Hangzhou 310058, China}
\affiliation{Institute of Fundamental and Transdisciplinary Research, and State Key Laboratory of Silicon and Advanced Semiconductor Materials, Zhejiang University, Hangzhou 310058, China}
\affiliation{Collaborative Innovation Center of Advanced Microstructures, Nanjing University, Nanjing, 210093, P. R. China}
	
\date{\today}
	
\begin{abstract}
   We report the crystal growth, structure, physical properties, and first-principles calculations of a vanadium-based oxytelluride Cs$_{1-\delta}$V$_2$Te$_2$O. The material possesses two-dimensional V$_2$O square nets sandwiched by tellurium layers, with local crystallographic symmetry satisfying the spin symmetry for a $d$-wave altermagnet. An antiferromagnetic transition at 293 K is unambiguously evidenced from the measurements of magnetic susceptibility and specific heat. In addition, a secondary transition at $\sim$70 K is also observed, possibly associated with a Lifshitz transition.
   The first-principles calculations indicate robust N\'{e}el-type collinear antiferromagnetism in the V$_2$O plane. Consequently, spin splittings show up in momentum space, in relation with the real-space mirror/rotation symmetry. Interestingly, the V-$d_{yz}/d_{xz}$ electrons, which primarily contribute the quasi-one-dimensional Fermi surface, turns out to be fully orbital- and spin-polarized, akin to the case of a half metal. Our work lays a solid foundation on the potential applications utilizing altermagnetic properties in vanadium-based oxychalcogenides.
\end{abstract}
	
\maketitle
	
\section{\label{sec:level1}Introduction}


Recently, a third form of magnetism, altermagnetism, was identified on the basis of spin-symmetry principles~\cite{2022PRX.edtorial,2022PRX.altermag}, the unconventionality of which can be traced back to the finding of an Fermi-surface instabilities in spin channel in 2007~\cite{2007PRB.WuCJ}. In an altermagnetic material, ordered spins are compensated by crystal symmetry of rotation or mirror (not an inversion), resulting in zero net magnetization yet with spin-split band structures in momentum space, as was earlier discovered independently by many groups~\cite{2019JPSJ.Hayami,2020PRB.Zunger,2020SA.Hall,2021NC.LiuJW,PNAS2021.Mazin}. Consequently, altermagnets may host an unusual combination of ferromagnetic (FM) and antiferromagnetic (AFM) properties, giving rise to responses of anomalous Hall~\cite{2020SA.Hall,2022NE.Hall,2024NC.Mn5Si3}, nonrelativistic spin-polarized current~\cite{2021NC.LiuJW}, tunneling magnetoresistance~\cite{2023PRB.TMR}, piezomagnetism~\cite{2021NC.LiuJW}, and even unitary triplet superconductivity~\cite{2022PRX.edtorial,2023PRB.SC,2023PRB.topoSC}, making an altermagnet highly potential for various applications.

While theory predicts that candidates of altermagnets can be common and diverse~\cite{2022PRX.landscape,Bai2024.AFM}, direct and solid experimental demonstration of altermagnets are not so many so far. 
In general, spin splitting in momentum space in an AFM state is considered as a hallmark for an altermagnet. In this context, only few materials such as MnTe~\cite{2024Nature.MnTe} and 
CrSb~\cite{2024NC.CrSbfilm,2024AS.CrSb.LiuC,2024PRL.CrSb.ShenDW,2025NC.CrSb.LiuY}, have been demonstrated to be an altermagnet by using spin- and angle-resolved photoemission spectroscopy. Different from these three-dimensional altermagnets above, very recently, quasi-two-dimensional altermagnetism was discovered in layered Rb$_{1-\delta}$V$_2$Te$_2$O~\cite{2025NP.Rb1221} and KV$_2$Se$_2$O~\cite{2025NP.K1221}.
These vanadium-based materials show a $d$-wave spin splitting with metallic conductivity, promising for realization of spin current to develop high-density, high-speed, and low-energy-consumption spintronic devices.
Nevertheless, little work has been performed on the intrinsic physical properties of this class of materials using single-crystalline samples, and the expected altermagnetic transition remains elusive~\cite{2024PRB.1221.ChenGF,2025NP.K1221}.

The vanadium-based materials mentioned above belong to the large family of `1221' oxychalcogenides $A$V$_2$$Ch_2$O ($A$ is an alkali metal; $Ch$ is a chalcogen)~\cite{Ablimit2018.thesis,2023JSSC.1221}. The very first member in this material family is the oxysulfide, CsV$_2$S$_2$O, which was identified as a paramagnetic bad metal~\cite{Valldor2015}. Two new members, the oxytelluride Rb$_{1-\delta}$V$_2$Te$_2$O~\cite{2018Ablimit} and the oxyselenide CsV$_2$Se$_2$O~\cite{2018PRB.WenHH}, were later reported in 2018. Notably, the alkali metal can be fully deintercalated, forming van der Waals compounds  V$_2$Se$_2$O~\cite{2018PRB.WenHH} and V$_2$Te$_2$O~\cite{2018IC.Ablimit}. Rb$_{1-\delta}$V$_2$Te$_2$O ($\delta\approx$ 0.2) exhibits a weak metal-to-metal transition at $\sim$100 K~\cite{2018Ablimit}, while the deintercalated V$_2$Te$_2$O is a correlated metal without any anomaly down to 30 mK~\cite{2018IC.Ablimit}. 
By contrast, the polycrystalline samples of the selenides CsV$_2$Se$_2$O and V$_2$Se$_2$O show a semiconducting behavior
~\cite{2018PRB.WenHH}. 
However, it was recently reported that single crystals of the selenide KV$_2$Se$_2$O show a metallic behavior with a density-wave (DW)-like transition at 105 K~\cite{2024PRB.1221.ChenGF}. These results underscore the necessity of physical property measurements using single-crystalline samples.



In this paper, high-quality crystals of a member of vanadium-based 1221 series, Cs$_{1-\delta}$V$_2$Te$_2$O ($\delta\approx$ 0.2), were successfully grown, and the intrinsic physical properties were systematically measured. An AFM transition at $T_\mathrm{N}=$ 293 K is observed unambiguously. Using density-functional-theory (DFT) calculations, we identify a N\'{e}el-type collinear antiferromagnetism in the V$_2$O plane as the ground state. 
The AFM metallic state shows unique orbital-selective momentum-dependent spin splittings at around Fermi energy, making the title compound potentially competitive in spintronic applications.
	
\section{\label{sec:level2}Experimental and calculation methods}

Single crystals of Cs$_{1-\delta}$V$_2$Te$_2$O were grown via a self-flux method~\cite{2018IC.Ablimit}. Mixtures of the source materials, Cs (99.8\%), V$_2$O$_5$ (99.99\%), Te (99.99\%), and V (99.9\%), in molar ratios of Cs~:~V~:~Te~:~V$_2$O$_5$ = 6.05~:~1.62~:~7~:~0.19 were loaded in a tubelike alumina crucible, then sealed in a Ta tube in Ar atmosphere by arc welding. This Ta tube was jacketed with a silica ampule. The sample-loaded assembly was heated slowly to 1000~$^{\circ}$C in a muffle furnace, held for 2000 min. Then it was allowed to cool to 960~$^{\circ}$C in 200 min., followed by slowly cooling at a rate of 2~$^{\circ}$C/h to 600~$^{\circ}$C before the furnace was switched off. Millimeter-sized platelike shiny crystals (Fig.~\ref{XRD}a) were mechanically separated from the flux in an argon-filled glove box. The crystals are sensitive to moisture, and exposure in ambient conditions should be avoided as far as possible.

The crystal structure was determined by single-crystal X-ray diffractions (XRD) using a Bruker D8 Venture diffractometer with Mo-$K_{\alpha}$ radiation. A piece of Cs$_{1-\delta}$V$_2$Te$_2$O crystal with dimensions of 0.05 $\times$ 0.02 $\times$ 0.02 mm$^3$ was mounted on the sample holder using oil of polybutenes. To confirm the structural inhomogeneity in large crystals that are suitable for the physical property measurements, we also carried out a $\theta-2\theta$ scan on a crystal plate (2 $\times$ 2 $\times$ 0.1 mm$^3$) using PANalytical X-ray diffractometer (Model EMPYREAN) with monochromatic Cu-$K_{\alpha1}$ radiation. The chemical composition of the as-grown crystals was measured using energy-dispersive x-ray spectroscopy (EDS) on a scanning electron microscope (Hitachi S-3700N) equipped with Oxford Instruments X-Max spectrometer.

The electrical resistivity, Hall coefficient, and heat capacity were measured on a Quantum Design Physical Property Measurement System. In the transport measurements, a rectangular thin crystal was chosen, on which electrodes were made using silver paste and gold wires. A standard six-electrode Hall bar method was employed, from which the longitudinal resistivity $\rho_{xx}$ and Hall resistivity $\rho_{xy}$ can be measured on the same sample. The heat capacity was measured with a relaxation method. The magnetization was measured on a Quantum Design Magnetic Property Measurement System with magnetic fields applied parallel and perpendicular to the crystal plate, respectively.

We also performed DFT-based first-principles calculations using VASP package~\cite{VASP}. The exchange-correlation energy was treated using generalized-gradient-approximation (GGA) functionals~\cite{GGA} with or without rotationally invariant on-site Coulomb interaction corrections ($U$)~\cite{LSDA+U}. The plane-wave energy cutoff for all calculations was set to 600 eV. We adopted 15$\times$15$\times$7 and 30$\times$30$\times$14 $\Gamma$-centered $\mathbf{k}$-mesh for self-consistent and density of states (DOS) calculations, respectively. The structure optimizations were performed in all magnetic calculations. The Fermi surface and orbital occupation were obtained from Wannier downfolding results~\cite{1997PRB.Wannier}. The band disentanglements were performed with initial projectors of ten V-3$d$, six Te-5$p$ and three O-2$p$ orbitals.

\section{\label{sec:level3}Results and discussion}

\subsection{\label{subsec:level1}Crystal structure}

\begin{figure}[htpb]
	\includegraphics[width=8cm]{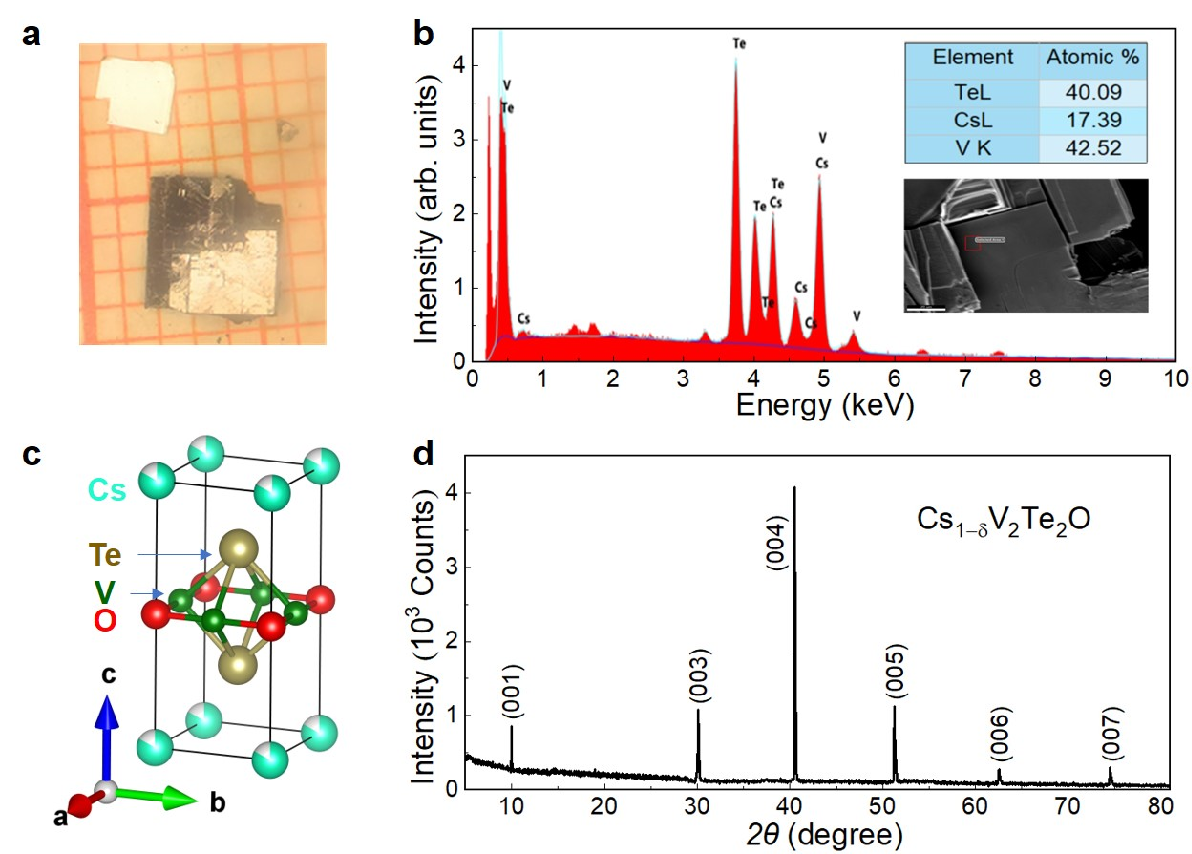}
	\caption{Characterizations of Cs$_{1-\delta}$V$_2$Te$_2$O ($\delta\approx$ 0.2) crystals by photography (a), EDS (b), crystal structure (c), and XRD (00$l$) diffractions at room temperature (d).}
	\label{XRD}
\end{figure}

The as-grown crystals of Cs$_{1-\delta}$V$_2$Te$_2$O were characterized by EDS and XRD. The EDS data (Fig.~\ref{XRD}b) confirms the chemical composition from Cs, V, Te, and O. The atomic ratios of Cs, V, and Te are typically 0.82~:~2.0~:~1.9 (oxygen content is not given here because of its large intrinsic measurement uncertainty). 
Consistently, the single-crystal XRD refinement yields a similar chemical formula Cs$_{0.83}$V$_2$Te$_2$O. Given the air sensitivity, which leads to a slight decrease in Cs content and, short-term exposure to ambient air is inevitable in experiments, we employ $\delta\approx$ 0.2 in the following description for sure. The crystallographic data of Cs$_{1-\delta}$V$_2$Te$_2$O at 250 K are presented in Table~\ref{structure}. The room-temperature (00$l$) diffractions by $\theta-2\theta$ scan are shown in Fig.~\ref{XRD}d, from which the $c$ axis is determined to be 8.909(2) \AA.
It is noted that, while the $c$ axis is $\sim$5.6\% larger than that (8.433 \AA) of Rb$_{0.83}$V$_2$Te$_2$O~\cite{2018Ablimit}, the $a$ axis is very close each other. The result indicates that the $a$ parameter is dominantly determined by the V$_2$Te$_2$O layers. 
The significant increase in $c$ axis suggests enhanced two dimensionality in Cs$_{1-\delta}$V$_2$Te$_2$O, which could suppress the magnetic transition temperature.

\begin{table}[htpb]
\caption{Crystallographic data of Cs$_{1-\delta}$V$_2$Te$_2$O from single-crystal X-ray diffraction at 250 K. The final $R$ index is $wR_2=$ 3.57\% , and the goodness of fit is $S=1.235$ for all data.}
  \label{structure}\renewcommand\arraystretch{1.3}
\begin{tabular}{p{1cm}<{\centering}|p{0.8cm}<{\centering}|p{0.8cm}<{\centering}
p{0.8cm}<{\centering}p{1.5cm}<{\centering}|p{1.2cm}<{\centering}|p{1.4cm}<{\centering}
}
      \hline \hline
    \multicolumn{3}{c}{Chemical Formula}   &\multicolumn{3}{c}{Cs$_{0.83}$V$_2$Te$_2$O}\\
\multicolumn{3}{c}{Space Group}  & \multicolumn{3}{c}{$P$4/$mmm$ (No. 123)} \\
\multicolumn{3}{c}{$a$ (\r{A})}   & \multicolumn{3}{c}{4.0430(6)} \\
\multicolumn{3}{c}{$c$ (\r{A})}   & \multicolumn{3}{c}{8.8533(19)} \\
    \hline
    Atom &  site  & $x$ & $y$ & $z$ &occ. &$U_{\mathrm{eq}}$ ({\AA}$^2$) \\
    \hline
    Cs   &  $1b$  & 0  & 0  & 0     &0.833(4)  & 0.0292(3) \\
    V   &  $2f$  & 0  & 1/2  & 1/2    &1       & 0.0112(2) \\
    Te    &  $2h$  & 1/2& 1/2  & 0.71780(5) &1  & 0.0131(2)\\
    O    &  $1a$  & 0 & 0 & 1/2             &1 & 0.0171(1) \\
\hline
    \hline
  \end{tabular}
\end{table}

The crystal structure of [V$_2$Te$_2$O]$^{1-\delta}$ layers host necessary symmetries for emergence of altermagnetism in the V sublattice. In a unit cell (Fig.~\ref{XRD}c), there are two V atoms, both of which are coordinated by two O$^{2-}$ and four Te$^{2-}$ ions, such that the two V atoms are connected by two diagonal mirror (vertical Te$_2$O planes) operations or four-fold rotation $C_{4z}$ (rather than inversion or any translation operations). In the case of a collinear N\'{e}el-type in-plane spin order (either G-type or C-type AFM), which is the case in the present material, the opposite-spin sublattices are connected by the spin group symmetry [$C_2\|M_{1\bar{1}0}$] or [$C_2\|C_{4z}$], where $C_2$ represents a spin-flip operation. This crystal-spin symmetry naturally leads to a momentum-dependent $d$-wave spin splittings (see below).

\subsection{\label{subsec:level2}Physical properties}

\begin{figure}[htpb]
	\includegraphics[width=8cm]{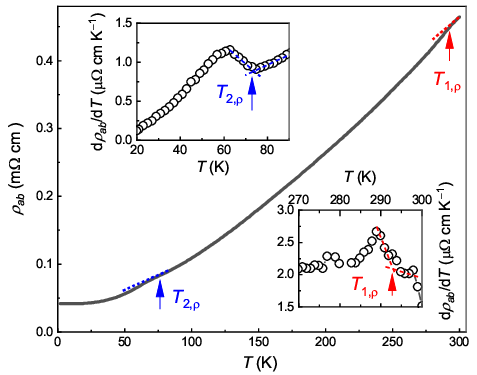}
	\caption{Temperature dependence of in-plane resistivity for Cs$_{1-\delta}$V$_2$Te$_2$O ($\delta\approx0.2$) crystals. Anomalies are indicated with arrows at $T_{1,\rho}\approx$ 292 K and $T_{2,\rho}\approx$ 70 K, respectively, which can also be seen in the derivative plots shown in the insets.}
	\label{Resistivity}
\end{figure}

Figure~\ref{Resistivity} shows the temperature dependence of in-plane resistivity, $\rho_{ab}(T)$, for the Cs$_{1-\delta}$V$_2$Te$_2$O ($\delta\approx0.2$) crystals. A metallic behavior is seen with an absolute room-temperature resistivity of about 0.5 m$\Omega$~cm and a residual resistivity ratio of $\rho_{\mathrm{300K}}/\rho_{\mathrm{2K}}\approx$ 11. Anomalies at $T_{1,\rho}\approx$ 292 K and $T_{2,\rho}\approx$ 70 K can be detected, respectively, from the first-order derivative of the resistivity curve. The magnetoresistance is about 0.5\% at around $T_{1,\rho}$ under a 9-T field (not shown here). Note that the behavior at $T_{2,\rho}$ resembles the weak metal-to-metal transition at 100 K in Rb$_{1-\delta}$V$_2$Te$_2$O~\cite{2018Ablimit} and the DW-like transition at 105 K in KV$_2$Se$_2$O~\cite{2024PRB.1221.ChenGF}. Notably, no additional anomaly was observed in the high-temperature regime up to 300 K for the latter two materials, probably because the transition temperatures are above room temperature~\cite{2025NP.Rb1221,2025NP.K1221}.

\begin{figure}[htpb]
	\includegraphics[width=8cm]{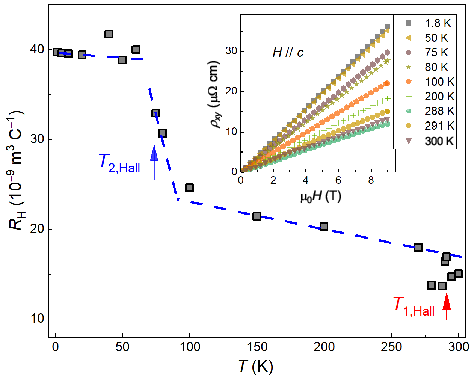}
	\caption{Temperature dependence of Hall coefficient of Cs$_{1-\delta}$V$_2$Te$_2$O ($\delta\approx0.2$) crystals. The error bars, derived from the linear fit of $\rho_{xy}(H)$, are smaller than the data-point size. The dashed blue lines are guides to the eye. Anomalies at $T_{1,\mathrm{Hall}}\approx$ 290 K and $T_{2,\mathrm{Hall}}\approx$ 75 K are marked, respectively. The inset shows the field dependence of Hall resistivity at representative temperatures for clarity.}
	\label{Hall}
\end{figure}

To further investigate the transport properties, we performed the Hall measurement. As shown in the inset of Fig.~\ref{Hall}, the Hall resistivity increases linearly with magnetic field, without detectable anomalous Hall effect for field along the $c$ axis. 
The positive values of the Hall coefficient, $R_\mathrm{H}=\rho_{xy}/(\mu_0H)$, indicate dominant hole-type charge transport. Assuming a single-band picture, one can estimate the hole concentration at room temperature, $n_\mathrm{h}=1/(eR_\mathrm{H})\sim4\times10^{26}$ m$^{-3}$, equivalent to a Hall number of 0.03 holes per vanadium atom. The estimated Hall number is much lower than the $3d$-electron counts of V$^{2.58+}$ (2.42). The result reversely suggests a multiband scenario, as supported by the DFT calculations (see below). The temperature dependence of $R_\mathrm{H}$ is displayed in Fig.~\ref{Hall}. One sees that the Hall coefficient increases abruptly at $T_{2,\mathrm{Hall}}\approx$ 75 K, suggesting decrease in the carrier concentration. Thus the corresponding resistivity kink at $T_{2,\rho}$ means an increase in carrier mobility ($\mu_{\mathrm{h},ab}~=~R_\mathrm{H}/\rho_{ab}$). Qualitatively similar results are observed in Rb$_{1-\delta}$V$_2$Te$_2$O~\cite{2018Ablimit} and KV$_2$Se$_2$O~\cite{2024PRB.1221.ChenGF}.
Subtle dip-like anomalies at around $T_{1,\mathrm{Hall}}\approx$ 290 K are also discernible, coincident with the AFM transition in the magnetic susceptibility measurements below.

\begin{figure}[htpb]
	\includegraphics[width=8cm]{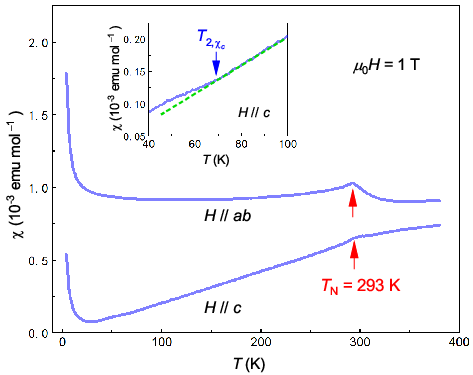}
	\caption{Temperature dependence of anisotropic magnetic susceptibility of Cs$_{1-\delta}$V$_2$Te$_2$O ($\delta\approx0.2$) crystals. 
The inset is a close-up for $\chi_{c}(T)$ where a small anomaly can be identified at $T_{2,\chi}\approx$ 70 K.}
	\label{MT}
\end{figure}

Figure~\ref{MT} shows the temperature dependence of anisotropic magnetic susceptibility for Cs$_{1-\delta}$V$_2$Te$_2$O ($\delta\approx0.2$). An AFM-like transition is clearly seen at $T_\mathrm{N}=$ 293 K with different behaviors in $\chi_{ab}(T)$ and $\chi_{c}(T)$. $\chi_{ab}(T)$ is basically temperature independent, except for a cusp at $T_\mathrm{N}$, whereas $\chi_{c}(T)$ decreases with decreasing temperature from 380 K down to $\sim$30 K, with a slope change at $T_\mathrm{N}$. The upturn tails at low temperatures correspond (through Curie-Weiss fit) to an effective magnetic moment of 0.14 and 0.11 $\mu_\mathrm{B}$/fu for $H\|ab$ and $H\|c$, respectively, which is too small to be attributed to an intrinsic local moment. The minor extrinsic moments likely come from lattice defects/imperfections rather than a specific secondary impurity phase, because the sample measured is a single crystal. Note that that $\chi_{c}$ is overall smaller than $\chi_{ab}$, similar to previous observations in Rb$_{1-\delta}$V$_2$Te$_2$O~\cite{2025NP.Rb1221} and KV$_2$Se$_2$O~\cite{2024PRB.1221.ChenGF,2025NP.K1221}, suggesting that spins are basically along the crystallographic $c$ axis. The unusual $\chi_{c}(T)$ above $T_\mathrm{N}$, which is neither Pauli paramagnetic nor Curie-Weiss paramagnetic, implies short-range AFM correlations or AFM fluctuations, reminiscence of the $\chi(T)$ behavior in the parent compounds of iron-based superconductors~\cite{2009EPL.ZhangGM}.

Compared with the magnetic susceptibility drop at the DW-like transition in Rb$_{1-\delta}$V$_2$Te$_2$O~\cite{2018Ablimit,2025NP.Rb1221} and KV$_2$Se$_2$O~\cite{2024PRB.1221.ChenGF}, the $\chi_{c}(T)$ anomaly at $\sim$70 K here is very weak (inset of Fig.~\ref{MT}) and, furthermore, no discernible anomaly can be seen in the $\chi_{ab}(T)$ data. The results basically rule out a CDW ordering (otherwise $\chi(T)$ would show a drop as usual).
Given the magnetic ordering is established already at 293 K, this transition should be secondary. 

\begin{figure}[htpb]
	\includegraphics[width=8cm]{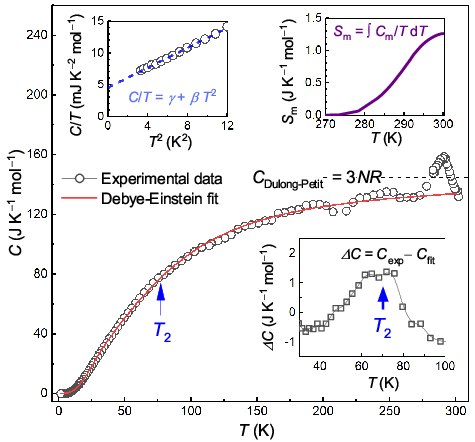}
	\caption{Temperature dependence of specific heat, $C(T)$, for Cs$_{1-\delta}$V$_2$Te$_2$O ($\delta\approx0.2$) crystals. 
The upper left plots $C$/$T$ as a function of $T^2$ in the low-temperature regime. The upper right shows the released magnetic entropy at the AFM transition. The lower right inset shows the difference between experimental and fitted data at around $T_2=$ 70 K. }
	\label{HC}
\end{figure}
	
Figure~\ref{HC} shows the specific heat data of Cs$_{1-\delta}$V$_2$Te$_2$O ($\delta\approx0.2$) crystals. A prominent peak is seen at 290 K, 
which obviously corresponds to the AFM transition evidenced from the magnetic susceptibility measurement above. The whole $C(T)$ data below the N\'{e}el temperature can be well fitted with a combined Debye-Einstein model~\cite{PhysRevB.64.012302}, which was extended to complex systems where independent Einstein oscillators are assumed to be embedded in a Debye framework composed of the heavier elements~\cite{2015PRB.2322}. The fitting formula follow:

\begin{flalign*}
C(T) = \gamma T + \xi C_E(\Theta_E, T)+(1-\xi) C_D(\Theta_D, T),
\end{flalign*}

where

\begin{flalign*}
C_E(\Theta_E, T) = 3NR\left(\frac{\Theta_E}{T}\right)^2 \frac{e^{\Theta_E/T}}{(e^{\Theta_E/T}-1)^2} ~\mathrm{and} \nonumber
\\
C_D(\Theta_D, T) =  9NR\left(\frac{T}{\Theta_D}\right)^3 \int_0^{\Theta_D/T} \frac{x^4 e^x}{(e^x-1)^2}  dx
\end{flalign*}

denote the Einstein and Debye terms, respectively. $\xi$ and ($1-\xi$) are their corresponding fractional weights, and $\Theta_E$ and $\Theta_D$ represents the characteristic Einstein and Debye temperatures. The first term $\gamma T$ is the electronic contribution, which can be accurately obtained by the analysis of low-temperature data below. As a result, the data fitting yields $\Theta_E = 321.1$ K, $\Theta_D = 150.6$ K, and $\xi = 0.50$. Notably, the fitted $\xi$ value is fully coincident with the fraction of light elements vanadium and oxygen, and $\Theta_E$ is indeed much higher than $\Theta_D$.

With fitted data above as the background, then, the magnetic contribution $C_{\rm m}$ can be obtained simply by a subtraction. The magnetic entropy released at the AFM ordering is estimated to be $S_{\rm m}$ = $\int$ $(C_{\rm m}/T)$d$T\approx$ 1.3 J K$^{-1}$ mol-fu$^{-1}$, which is an order of magnitude smaller than the expected one: 2$R$ln($2S+1$) = 11.5 J K$^{-1}$ mol-fu$^{-1}$ even with the smallest value of $S=$1/2 (the prefactor 2 comes from two V atoms in one formula unit). The result suggests existence of short-range spin correlations above $T_\mathrm{N}$, consistent with the unusual $\chi_{c}(T)$ behavior. 

The subtraction of $C(T)$ data also uncover an anomaly at $T_2\approx$ 70 K, as shown in the lower inset of Fig.~\ref{HC}. The magnitude of the anomaly is quite similar to those of Rb$_{1-\delta}$V$_2$Te$_2$O~\cite{2018Ablimit} and KV$_2$Se$_2$O~\cite{2024PRB.1221.ChenGF}, suggesting a common origin. In KV$_2$Se$_2$O, an SDW transition is concluded from the nuclear magnetic resonance experiment~\cite{2025NP.K1221}, while a change in the vanadium valence state is suggested by the electron energy loss spectroscopy~\cite{2025EPL.Zhuang}. Very recently, a theoretical investigation~\cite{2025PRB.K1221.DMFT} suggests a Lifshitz transition with the disappearance of electron-like Fermi pockets near the $\Gamma$ point, for the similar anomaly in KV$_2$Se$_2$O. Given the much weaker anomaly at $\sim$70 K here in Cs$_{1-\delta}$V$_2$Te$_2$O, especially for the magnetic and heat-capacity measurements, such an electronic transition is very likely. Notably, if it is the case, the Lifshitz transition does not alter the altermagnetism. 

On the upper left of Fig.~\ref{HC} we plot the low-temperature data in $C/T$ versus $T^2$. From the linear fit, a Sommerfeld coefficient of $\gamma=$ 4.5 mJ K$^{-2}$ mol$^{-1}$ and a Debye coefficient of $\beta=$ 0.785 mJ K$^{-4}$ mol$^{-1}$ are obtained. The $\gamma$ value is lower than that of Rb$_{1-\delta}$V$_2$Te$_2$O (7.5 mJ K$^{-2}$ mol$^{-1}$)~\cite{2018Ablimit}, yet larger than that of KV$_2$Se$_2$O (1.9 mJ K$^{-2}$ mol$^{-1}$)~\cite{2024PRB.1221.ChenGF}. Our DFT calculations below show that the theoretical values are $\gamma_0=\frac{1}{3}\pi^2k_\mathrm{B}^2N(E_\mathrm{F})=$ 5.0 and 9.8 mJ K$^{-2}$ mol$^{-1}$, respectively, for the AFM ground state [$N(E_\mathrm{F})=$ 2.13 eV$^{-1}$ fu$^{-1}$ for both spins] and for the non-magnetic state [$N(E_\mathrm{F})=$ 4.14 eV$^{-1}$ fu$^{-1}$]. The result supports that realization of the AFM state, and suggests weak electron correlations in the system. 
From the fitted $\beta$ value, the Debye temperature is estimated to be 243 K using the formula $\Theta_D=(\frac{12}{5}\pi^4NR/\beta)^{1/3}$, where $N=$ 5.8 (the number of atoms per formula unit) and $R$ is the gas constant. One notes that this Debye temperature is nearly the average of $\Theta_E$ and $\Theta_D$ derived above within the combined Debye-Einstein model. The apparent discrepancy on Debye temperature stems from the fundamentally different physical models upon which they are based.

From the systematic measurements of electrical transport, magnetic susceptibility, and specific heat on the Cs$_{1-\delta}$V$_2$Te$_2$O  ($\delta\approx0.2$) crystals above, one concludes a primary AFM transition at 293 K. The N\'{e}el temperature is lower than those of its sister compounds Rb$_{1-\delta}$V$_2$Te$_2$O (307 K)~\cite{2025NP.Rb1221} and KV$_2$Se$_2$O ($>$300 K)~\cite{2025NP.K1221}. Given a similar $\delta$ value, a similar lattice parameter $a$, yet a $\sim$5\% larger $c$ axis for Cs$_{1-\delta}$V$_2$Te$_2$O, compared with Rb$_{1-\delta}$V$_2$Te$_2$O, the reduced $T_\mathrm{N}$ here is possibly due to enhanced two dimensionality.

\subsection{\label{subsec:level3}First-principles calculations}

\begin{figure*}[htpb]
	\includegraphics[width=14cm]{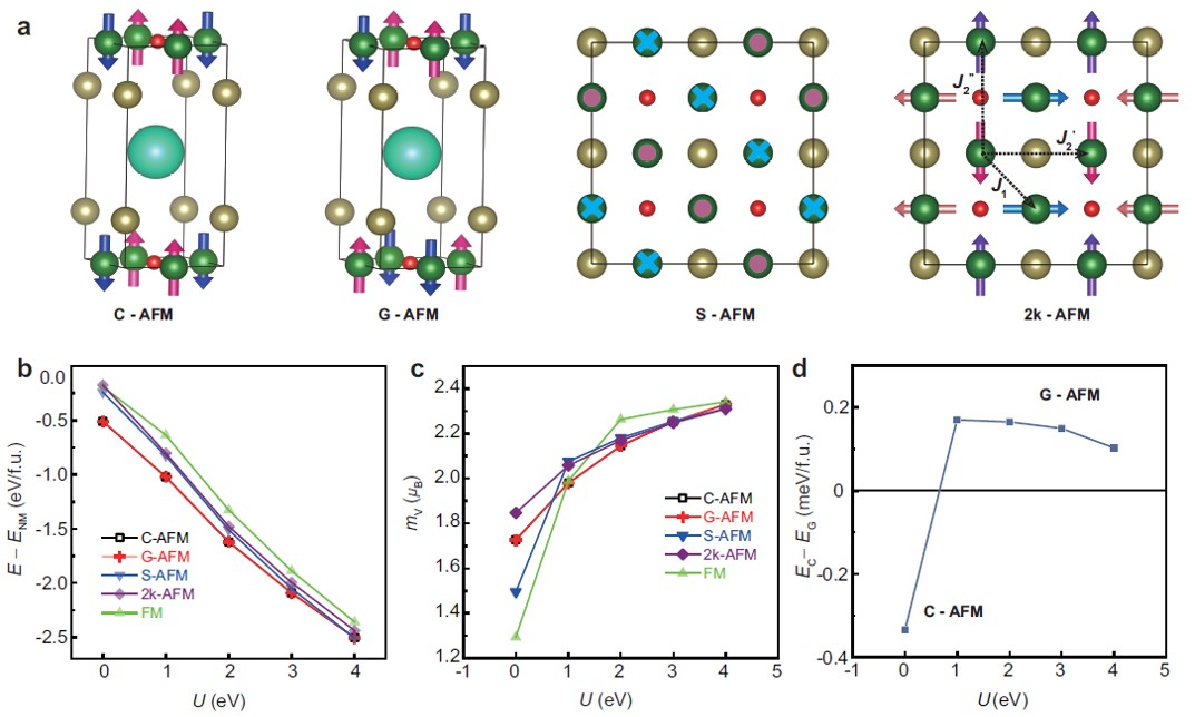}
	\caption{(a) Possible magnetic orders in CsV$_2$Te$_2$O: C-, G-, S-, and 2$\mathbf{k}$-antiferromagnetism (AFM), in addition to ferromagnetism (FM). The vanadium spins are aligned along the $c$-axis, except for the 2$\mathbf{k}$-AFM where spins lie in the $a$- or $b$-direction. For the S- and 2$\mathbf{k}$-AFM, ferromagnetic spin order along the $c$ axis is considered for simplicity. The intralayer magnetic exchange couplings, $J_1$, $J_2'$, and $J_2''$, are marked in the 2$\mathbf{k}$-AFM panel. (b-d) Magnetic energy, magnetic moment, and magnetic energy difference between C- and G-AFM as functions of $U$ for the different magnetic states of CsV$_2$Te$_2$O.}
\label{MS}
\end{figure*}

To understand the physical properties and to explore the underlying altermagnetism in CsV$_2$Te$_2$O, we carried out the DFT-based calculations. Let us first investigate the possible magnetic ground state. Considering the two-dimensional (2D) crystal structure with weak out-of-plane magnetic coupling, one may first evaluate the dominant in-plane magnetic configurations. In a 2D square lattice with short-range magnetic interactions--nearest-neighbor (NN) and next-nearest-neighbor (NNN) couplings, there are three possible collinear magnetic orders: namely, N\'{e}el-type AFM, striped AFM, and FM. So, we calculate the magnetic energies relative to the energy of non-magnetic (NM) configuration for the C-, G-, S-type AFM states with spins along the $c$ axis [see Fig.~\ref{MS}a] as well as the FM state in the GGA$+U$ scheme. In addition, we also consider a non-collinear magnetic order with spins lying in the $ab$ plane, called 2$\mathbf{k}$-AFM, which often appears in the analogous structure~\cite{2016Stock}. 

Figure~\ref{MS}b shows the magnetic energies of the five magnetically ordered states as functions of $U$. The magnetic energies are all negative, and decrease with increasing $U$, indicating robust magnetic states over the NM one. The C- and G-type orders, both of which possess in-plane N\'{e}el-type AFM, turn out to be the ground state for $U\leq$ 4 eV. Notably, these two AFM configurations almost have the same energy, consistent with insignificantly small out-of-plane magnetic coupling. 
It seems that the C-type AFM slightly wins at $U=$ 0, yet the G-type AFM becomes more stable at $U\geq$ 1 eV, for spins along the $c$ axis (Fig.~\ref{MS}d). 
Note that the energy differences are minute, and they could be influenced by the Cs deficiency and other factors. The lowest magnetic moment is 1.3 $\mu_\mathrm{B}/\mathrm{V}$ at $U=0$ eV, which increases with $U$, and converges to $\sim$2.3 $\mu_\mathrm{B}/\mathrm{V}$ at the large $U$. The latter value is close to the spin-only one (2.5 $\mu_\mathrm{B}/\mathrm{V}$) at free-ion (V$^{2.5+}$ with $3d^{2.5}$) limit. While the C-AFM  turns out to be the most stable magnetic configuration for CsV$_2$Te$_2$O at $U=0$ eV, the real magnetic structure for Cs$_{1-\delta}$V$_2$Te$_2$O ($\delta\approx0.2$) needs to be experimentally determined by neutron diffractions and other experiments in the future.

\begin{table*}[htpb]
\caption{Magnetic energy $E_\mathrm{m}$ (meV/f.u.) and magnetic moment $\mu_{\rm V}$ ($\mu_\mathrm{B}$) of five different magnetically ordered states (see Fig.~\ref{MS}a) of CsV$_2$Te$_2$O in the GGA+$U$ calculations. For S-AFM and 2$\mathbf{k}$-AFM, the ferromagnetic interplane orders are considered. $J_1$, $J_2'$, and $J_2''$ denote in-plane spin-exchange parameters derived within $J_1-J_2'-J_2''$ Heisenberg model (see the text for details). }
	\renewcommand\arraystretch{1.3}
\begin{tabular}{p{2cm}<{\centering}|p{1.4cm}<{\centering}p{1.4cm}<{\centering}|p{1.4cm}<{\centering}
p{1.4cm}<{\centering}|p{1.4cm}<{\centering}p{1.4cm}<{\centering}|p{1.4cm}<{\centering}
p{1.4cm}<{\centering}|p{1.4cm}<{\centering}p{1.4cm}<{\centering}}
\hline
\hline
\multirow{2}{*}{\begin{tabular}[c]{@{}c@{}}$U$ (eV)\end{tabular}} & \multicolumn{2}{c|}{0} & \multicolumn{2}{c|}{1} & \multicolumn{2}{c|}{2} & \multicolumn{2}{c|}{3} & \multicolumn{2}{c}{4} \\ \cline{2-11}
& $E_{\mathrm{m}}$ & $\mu_{\rm V}$ & $E_{\mathrm{m}}$  & $\mu_{\rm V}$  & $E_{\mathrm{m}}$        & $\mu_{\rm V}$ & $E_{\mathrm{m}}$  & $\mu_{\rm V}$ & $E_{\mathrm{m}}$ & $\mu_{\rm V}$     \\
\hline
 FM          & $-$189.14 & 1.294 & $-$635.81 & 1.990 & $-$1327.37 & 2.263 & $-$1887.90 & 2.306 & $-$2359.09 & 2.341 \\
    C-AFM       & $-$509.78 & 1.727 & $-$1020.46 & 1.980 & $-$1621.29 & 2.145 & $-$2091.50 & 2.252 & $-$2504.36 & 2.329 \\
    G-AFM       & $-$509.44 & 1.726 & $-$1020.63 & 1.980 & $-$1621.46 & 2.145 & $-$2091.65 & 2.252 & $-$2504.47 & 2.329 \\
    S-AFM       & $-$241.95 & 1.496 & $-$825.43 & 2.076 & $-$1512.10 & 2.180 & $-$2050.04 & 2.254 & $-$2508.28 & 2.310 \\
    2$\mathbf{k}$-AFM & $-$174.15 & 1.847 & $-$802.69 & 2.058 & $-$1478.03 & 2.169 & $-$1995.74 & 2.247 & $-$2439.28 & 2.309 \\
\hline
 $J_1S^2$(\si{meV})    & \multicolumn{2}{c|}{40.1} & \multicolumn{2}{c|}{48.1} & \multicolumn{2}{c|}{36.7} & \multicolumn{2}{c|}{25.4} & \multicolumn{2}{c}{18.2} \\
 $J_2'S^2$(\si{meV})       & \multicolumn{2}{c|}{17.0} & \multicolumn{2}{c|}{5.7} & \multicolumn{2}{c|}{8.5} & \multicolumn{2}{c|}{13.6} & \multicolumn{2}{c}{17.2} \\
 $J_{2}^{''}S^2$(\si{meV}) & \multicolumn{2}{c|}{$-$43.8} & \multicolumn{2}{c|}{$-$6.4} & \multicolumn{2}{c|}{0.9} & \multicolumn{2}{c|}{1.5} & \multicolumn{2}{c}{1.9} \\
\hline
\hline
\end{tabular}
	\label{magnetic structure}
\end{table*}

\begin{figure*}[htpb]
	\includegraphics[width=16cm]{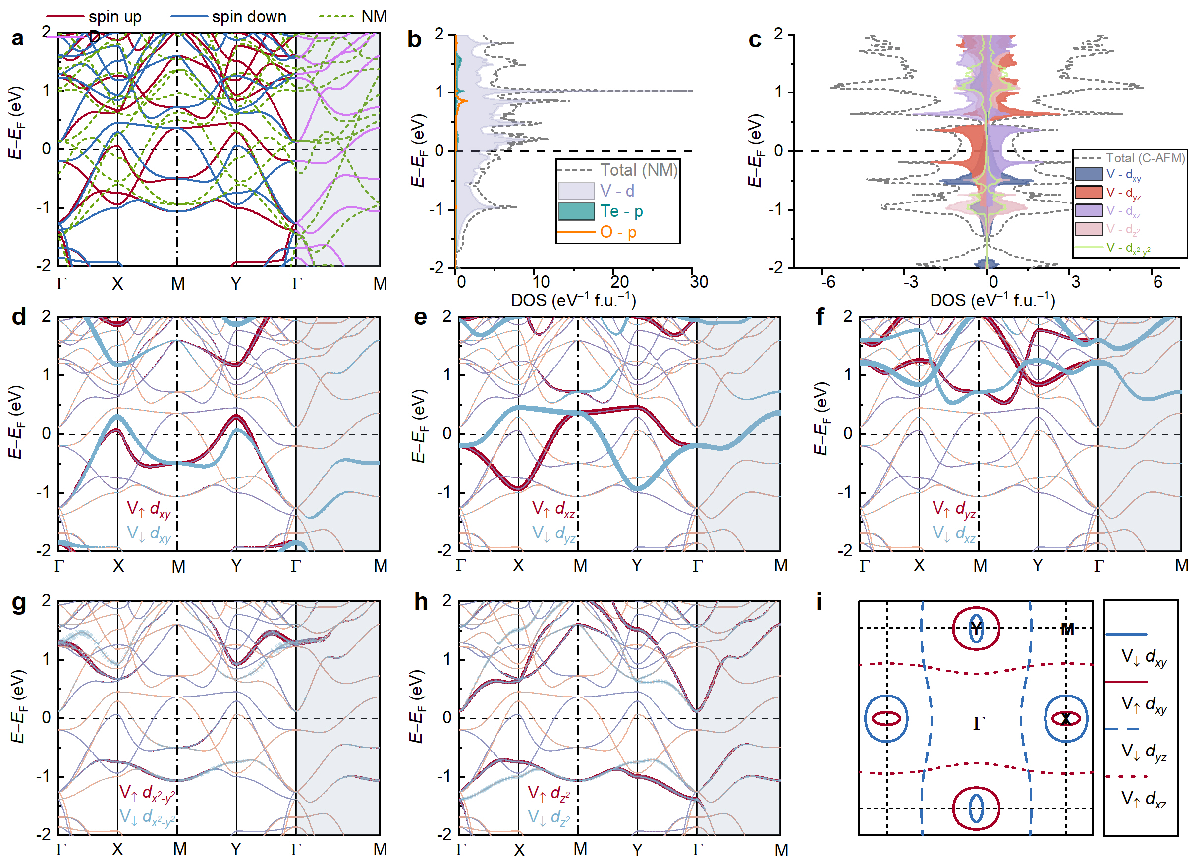}
	\caption{Band structure (a) and density of states projected on each atom (b) and each vanadium 3$d$ orbital (c) for C-AFM and non-magnetic (NM) CsV$_2$Te$_2$O. Panels (d-h) highlight the contributions from the five V-3$d$ orbitals. (i) The Fermi-surface sheets in the $k_z=0$ plane dominated by V- $d_{xy}$ and $d_{xz/yz}$ electrons.}
	\label{BS}
\end{figure*}

Now let us investigate on the magnetic interactions among the V-ion spins. 
For the specific coordination of vanadium ions, the effective magnetic interactions can be modelled by, i) NN coupling $J_1$ mainly through direct exchange, $\sim$60$^{\circ}$ V$-$Te$-$V superexchange (SE), and 90$^{\circ}$ V$-$O$-$V SE interactions, ii) NNN coupling $J_2'$ with $\sim$90$^{\circ}$ V$-$Te$-$V SE interaction, and iii) another NNN coupling $J_2''$ with 180$^{\circ}$ V$-$O$-$V SE and double-exchange (DE) interactions.
Then, the magnetic energies of the four magnetic structures shown in Fig.~\ref{MS}a can be expressed in terms of the $J_1-J_2'-J_2''$ model with the Heisenberg Hamiltonian, $H=\sum_{<i,j>} J_{ij}\mathbf{S_{i}}\cdot \mathbf{S_{j}}$,

\begin{subequations}
\begin{eqnarray}
E_{\mathrm{FM}}-E_{0}=(4J_1+2J_2'+2J_2'')S^2, \label{1a}\\
E_{\mathrm{S}}-E_{0}=(-2J_2'-2J_2'')S^2, \label{1b}\\
E_{\mathrm{C}}-E_{0}=(-4J_1+2J_2'+2J_2'')S^2, \label{1c}\\
E_{\mathrm{2k}}-E_{0}=(2J_2'-2J_2'')S^2, \label{1d}
\end{eqnarray}
\end{subequations}
where $E_{\mathrm{FM}}$, $E_{\mathrm{S}}$, $E_{\mathrm{C}}$, and $E_{\mathrm{2k}}$ denote their magnetic energy, respectively, and $E_0$ is the reference energy. The obtained spin-exchange parameters, $J_1S^2$, $J_2'S^2$, and $J_2''S^2$, are presented in Table~\ref{magnetic structure}. The results show that $J_1$ is mostly dominant, and its positive sign (antiferromagnetic coupling) dictates the in-plane N\'{e}el-type order in the magnetic ground state. The decrease in $J_1$ with increasing the $U$ parameter can be explained by the relationship, $J_1 \sim J_{\mathrm{SE}} \sim t^2/U_0$, where $t$ is the hopping integral and $U_0$ is the onsite Coulomb interaction. The exceptionally large ferromagnetic  $J_2''$ at $U=$ 0 could be due to Zener DE interaction between V$^{3+}$ and V$^{2+}$ in the itinerant-electron regime. For a large $U$, however, the conduction electrons tend to localize, favoring a $\sqrt{2}\times\sqrt{2}$ charge order in which the SE interactions of V$^{2+}-$O$-$V$^{2+}$ and V$^{3+}-$O$-$V$^{3+}$ are anticipated to significantly contribute an AFM interaction. Consequently, the SE and DE interactions largely cancel each other out, resulting in a small $J_2''$ value at a large $U$. 

Figures~\ref{BS}a-c show the band structure and DOS projected on each atom and on each vanadium 3$d$ orbital for the C-AFM and NM CsV$_2$Te$_2$O. Due to the 2D nature in the electronic structure of the analogous materials~\cite{2025NP.K1221,2025NP.Rb1221}, the band structure is presented only in the $k_z=$ 0 plane. Compared with the NM state, the band structure of the C-AFM state changes significantly nearby the Fermi level ($E_\mathrm{F}$). In particular, spin splitting is evident along $\Gamma-$X(Y)$-$M lines (left side in Fig.~\ref{BS}a). By contrast, no spin splitting (spin degenerate in other word) is seen along the $\Gamma-$M line (shaded area). The result demonstrates a $d$-wave altermagnetism in CsV$_2$Te$_2$O, because the mirror ($\Gamma-$M line in the mirror plane) symmetry connects the opposite spins in momentum space, directly corresponding to the real-space spin group symmetry of [$C_2\|M_{1\bar{1}0}$] that connects upper and down spins. Note that the DOS at around $E_\mathrm{F}$ are predominantly contributed by vanadium 3$d$ electrons for both NM and C-AFM states, yet the DOS for C-AFM is significantly reduced. Also, the opposite spins are fully compensated in the C-AFM state, in accordance with the spin-symmetry principle. Here we note that, in the case of the G-AFM state, our calculations indicate (not shown here) that each V$_2$O layer exhibits the same altermagnetism, yet the neighboring V$_2$O layer cancel the altermagnetism out. This phenomenon has recently been given the name `hidden altermagnetism'~\cite{2025PRL.hidden-AM,2025FP.Hidden-AM}.

Previously, we found that, in RbV$_2$Te$_2$O, there is an obvious orbital polarization of vanadium $d_{xz}$ and $d_{yz}$ electrons at around $E_\mathrm{F}$~\cite{2018Ablimit}. To investigate its relevance to altermagnetic order, we carried out V-3$d$-orbital projections onto the band structure. As shown in Figs.~\ref{BS}e,f, the momentum-dependent spin splittings exactly correspond to the V-$d_{xz}$/$d_{yz}$ orbital polarization: $d_{xz}$ and $d_{yz}$ electrons `split' along $\Gamma-$X(Y)$-$M lines, whereas they are orbitally degenerate along the $\Gamma-$M line. The result suggests a `spin-orbital locking', which can be understood in terms of Hubbard model qualitatively. 
The spin-up $d_{xz}$ (spin-down $d_{yz}$) electrons have energies $\sim$1.5 eV lower than the spin-down $d_{xz}$ (spin-up $d_{yz}$) electronic states. In this context, the spin-up $d_{xz}$ (spin-down $d_{yz}$) states can be viewed as in the lower Hubbard bands, whereas the spin-down $d_{xz}$ (spin-up $d_{yz}$) states serve as the upper Hubbard bands. There is a `gap' between the lower and upper Hubbard bands (Figs.~\ref{BS}e,f), leading to a full orbital/spin polarization of $d_{xz}$ and $d_{yz}$ electrons. Namely, the V- $d_{xz}$ and $d_{yz}$ electrons behave as a half metal. The spin-up $d_{xz}$ and the spin-down $d_{yz}$ bands forms quasi-one-dimensional Fermi surface (FS) sheets along $k_x$ and $k_y$ directions, respectively (Fig.~\ref{BS}i). In comparison, the $d_{xy}$ bands is basically isotropic in the $ab$-plane, forming $d$-wave-like spin-polarized FS pockets. 
The other two V-3$d$ orbitals, $d_{x^2-y^2}$ and $d_{z^2}$, contribute little at $E_\mathrm{F}$, as is see in Figs.~\ref{BS}c,g,h, indicating an orbital-selective altermagnetism in the present system. 

\

\section{\label{sec:level4}Concluding Remarks}

In summary, high-quality single crystals of a vanadium-based oxychalcogenide, Cs$_{1-\delta}$V$_2$Te$_2$O ($\delta\approx0.2$), were grown, on which the intrinsic physical properties were measured. The experimental results show a primary AFM ordering at 293 K. 
DFT calculations suggest a collinear in-plane N\'{e}el-type AFM ground state, which, by spin-symmetry principles, gives rise to momentum-dependent $d$-wave spin splittings in the electronic bands. The DOS at $E_\mathrm{F}$ is dominated by V-$t_{2g}$ electrons. Interestingly, there exists a spin-orbital locking for the $d_{xz}$ and $d_{yz}$ electrons, manifested by a concurrent full orbital/spin polarization at around $E_\mathrm{F}$. The fully spin- orbital-polarized electrons constitute the quasi-one-dimensional FS.

Materials with tetragonal $Tm_2Ch_2$O ($Tm$ and $Ch$ denote a transition metal and a chalcogen element, respectively) layers, or simply the $Tm_2Ch_2$O monolayers, possess necessary crystal symmetry for $d$-wave altermagnetism. In the case of a collinear N\'{e}el-type in-plane magnetic order for the $Tm$-ion spins, spin splittings in momentum space naturally appear without spin-orbit coupling, as has been studied in many related theoretical works~\cite{2023PRB.221.TMR,2024APL.221,2024NL.Fe2Se2O,2025JPCM.221,2025PRB.V2Te2O}. Nevertheless, the experimental realization is very limited so far. Comparatively, vanadium-based systems show robust collinear N\'{e}el-type in-plane spin order with quasi-two-dimensional conductivity, beneficial for practical applications. This work demonstrates an additional vanadium-based altermagnet candidate with the N\'{e}el temperature at room temperature. Given the variations in the N\'{e}el temperature within the vanadium-based 1221 family~\cite{2025NP.Rb1221,2025NP.K1221}, our result here suggests tunability of the N\'{e}el temperature, which is worthy of further investigations. 
\

\begin{acknowledgments}
We thank Junwei Liu, Chaoyu Chen, Yue Zhao, and Yang Liu for helpful discussions. This work was supported by the National Key Research and Development Program of China (2022YFA1403202 and 2023YFA1406101).
\end{acknowledgments}

\bibliography{1221}

\end{document}